\documentclass[journal=nalefd,manuscript=article]{achemso}

\usepackage[version=3]{mhchem}
\usepackage{enumitem}

\title{Nonlinear anisotropic dielectric metasurfaces for ultrafast nanophotonics}

\author{Giuseppe~Della~Valle}
\affiliation{Dipartimento di Fisica, Politecnico di Milano, and Istituto di Fotonica e Nanotecnologie del Consiglio Nazionale delle Ricerche, Piazza Leonardo da Vinci 32, 20133 Milano, Italy}
\email{giuseppe.dellavalle@polimi.it}

\author{Ben~Hopkins}
\affiliation{Nonlinear Physics Centre, Research School of Physics and Engineering, The Australian National University, Canberra, ACT 2601, Australia}

\author{Lucia~Ganzer}
\author{Tatjana~Stoll}
\affiliation{Dipartimento di Fisica, Politecnico di Milano, and Istituto di Fotonica e Nanotecnologie del Consiglio Nazionale delle Ricerche, Piazza Leonardo da Vinci 32, 20133 Milano, Italy}

\author{Mohsen~Rahmani}
\affiliation{Nonlinear Physics Centre, Research School of Physics and Engineering, The Australian National University, Canberra, ACT 2601, Australia}

\author{Stefano~Longhi} 
\affiliation{Dipartimento di Fisica, Politecnico di Milano, and Istituto di Fotonica e Nanotecnologie del Consiglio Nazionale delle Ricerche, Piazza Leonardo da Vinci 32, 20133 Milano, Italy}

\author{Yuri~S.~Kivshar}
\affiliation{Nonlinear Physics Centre, Research School of Physics and Engineering, The Australian National University, Canberra, ACT 2601, Australia}

\author{Costantino~De~Angelis}
\affiliation{CNISM and Dipartimento di Ingegneria dell'Informazione, Universit\'a di Brescia, Via Branze 38, 25123 Brescia, Italy}

\author{Dragomir~N.~Neshev}
\affiliation{Nonlinear Physics Centre, Research School of Physics and Engineering, The Australian National University, Canberra, ACT 2601, Australia}
\email{Dragomir.Neshev@anu.edu.au}

\author{Giulio~Cerullo}
\affiliation{Dipartimento di Fisica, Politecnico di Milano, and Istituto di Fotonica e Nanotecnologie del Consiglio Nazionale delle Ricerche, Piazza Leonardo da Vinci 32, 20133 Milano, Italy}

\begin{document}

\newpage
\begin{abstract}
We report on the broadband transient optical response from anisotropic nanobrick amorphous silicon particles, exhibiting Mie-type resonances. A quantitative model is developed to identify and disentangle the three physical processes that govern the ultrafast changes of the nanobrick optical properties, namely two-photon absorption, free-carrier relaxation, and lattice heating. We reveal a set of operating windows where ultrafast all-optical modulation of transmission is achieved with full return to zero in 20~ps. This is made possible due to the interplay between the competing nonlinear processes and despite the slow (nanosecond) internal lattice dynamics. The observed ultrafast switching behavior can be independently engineered for both orthogonal polarizations using the large anisotropy of nanobricks thus allowing ultrafast anisotropy control. Our results categorically ascertain the potential of all-dielectric resonant nanophotonics as a platform for ultrafast optical devices, and reveal the possibility for ultrafast polarization-multiplexed displays and polarization rotators.

\end{abstract}

\maketitle

\begin{center}
{\bf TOC Graphics}
\begin{figure}
\includegraphics[width=8.5cm]{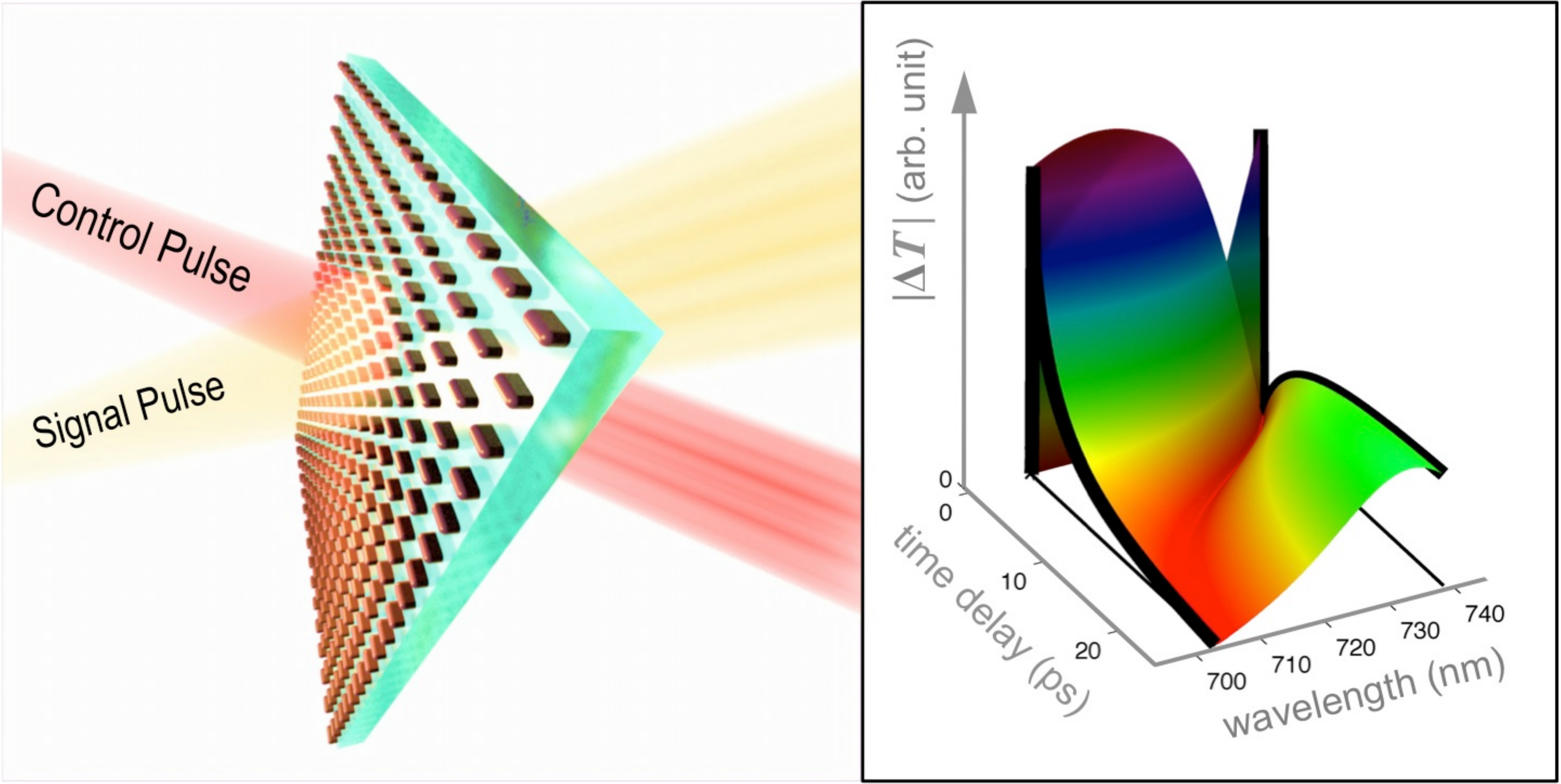}
\end{figure}
\end{center}

{\bf Keywords:} {\it Mie resonances, dielectric metasurfaces, Nonlinear optics, ultrafast spectroscopy, all-optical modulation}

\newpage

Following the growth of nanofabrication technologies, there has been a burgeoning interest in periodic arrangements of resonant nanostructures, tightly packed as {\it meta-atoms} that can form synthetic two-dimensional materials, or {\it metasurfaces}. These have enabled a host of novel applications for flat optics~\cite{Kildishev_SCI_2013, Capasso_NatMat_2014} and are now poised also toward nonlinear optical functionality~\cite{Guixin:NRM:2017}. The first meta-atoms utilized plasmonic resonances in noble metals, which also possess a strong optical nonlinearity~\cite{Sun_PRB_1994, Boyd_OC_2014}. The emergent vision was to exploit the nonlinearity in such plasmonic meta-atoms as a new route for all-optical modulation and switching, and, more generally, for ultrafast nanophotonics~\cite{Wurtz_NN_2011, Kauranen_NP_2012, Baida_PRL_2012, Brinks_PNAS_2013, Zavelani_ACSP_2015, Wang_NL_2015, Harutyunyan_NN_2015, Faggiani_ACSP_2017, Ciappina_RPP_2017, Stockman_NJP_2008}. However, the most eligible plasmonic metals exhibit very high linear and nonlinear ohmic losses, intrinsically related to the localized fields in a plasmonic resonance~\cite{Khurgin_MRS_2012} and are only weakly compatible with the large scale CMOS integration platform~\cite{Naik_AM_2013}. The search for better plasmonic materials with lower losses remains a current and growing topic of interest~\cite{Boltasseva_Science_2011}, where heavily-doped semiconductors~\cite{Comin_CSR_2014, Scotognella_EPJB_2013} and graphene~\cite{Koppens_NL_2011} are good new candidates.

{\sl All-dielectric} nano-resonators present an alternate route for nonlinear nanophotonics by exploiting Mie-type resonances instead of plasmonics. High refractive index dielectrics offer high quality, localized resonances that can enable access to nonlinear functionalities while maintaining minimal ohmic losses~\cite{Kuznetsov16:Sci,Liu_PQE_2017}. Silicon has been the material of choice for nonlinear dielectric nano-resonators~\cite{Shcherbakov:2014:NL,Yang:2015:NL, Shorokhov:2016:NL} offering strong nonlinear response and two-photon absorption (TPA)~\cite{Ikeda_OE_2007}. Proof of principle attosecond experiments on silicon-based dielectrics also indicate the viability of extreme switching speeds with a bandwidth up to the petahertz~\cite{Krausz_Nat_2016,Vampa_NP_2017}.

Recently, all-optical switching in planar array of a-Si:H nano-resonators has also been demonstrated, exploiting Mie-like magnetic~\cite{Shcherbakov_NL_2015, Baranov_ACSP_2016} and Fano resonances~\citep{Yang:2015:NL}. The observed all-optical modulation was found to be a result of the combined action of instantaneous TPA, free carrier generation and thermal effects. However, despite the observed ultrafast dynamics, only modulation at a single wavelength and a single polarization has been measured, thus missing a plethora of opportunities for ultrafast spectral and polarization control. Furthermore, the interplay of the underlying physical mechanisms for optical modulation was never addressed, leaving it unclear as to what control is available over the transient dynamics.

Even more concerning is that the TPA and the free-carrier relaxation are inevitably accompanied by lattice heating~\cite{Fauchet_JNCS_1992}, which in turn contributes to the optical modulation via the thermo-optic effect. This contribution is long-lived, because the cooling of the lattice is governed by slow phonon-phonon scattering processes that have nanosecond relaxation times, resulting in an inherent limitation to the switching speed of nanoscale silicon resonators for nonlinear nanophotonics. Here, we present an avenue to overcome such limitations and then demonstrate ultrafast all-optical transmission modulation in a-Si:H with full recovery on the picosecond timescale. Our approach starts from the experimental and theoretical analyses of the large optical nonlinearity exhibited by a-Si:H metasurfaces with anisotropic meta-atoms across the whole visible spectrum. Such combined study enables to disentangle the physical mechanisms governing the observed nonlinearity, and quantitatively elucidate the contribution of each to the transient modulation of the anisotropic Mie-like resonances. This allows us to identify the avenue by which slow processes that limit modulation speed can be suppressed by superimposing different nonlinear mechanisms taking place within a given resonance, or even  competing effects from two neighboring resonances. These results offer a new premise for exploiting all-dielectric metasurfaces for ultrafast spectral and polarization switching.

\begin{figure}
\includegraphics[width=16.5cm]{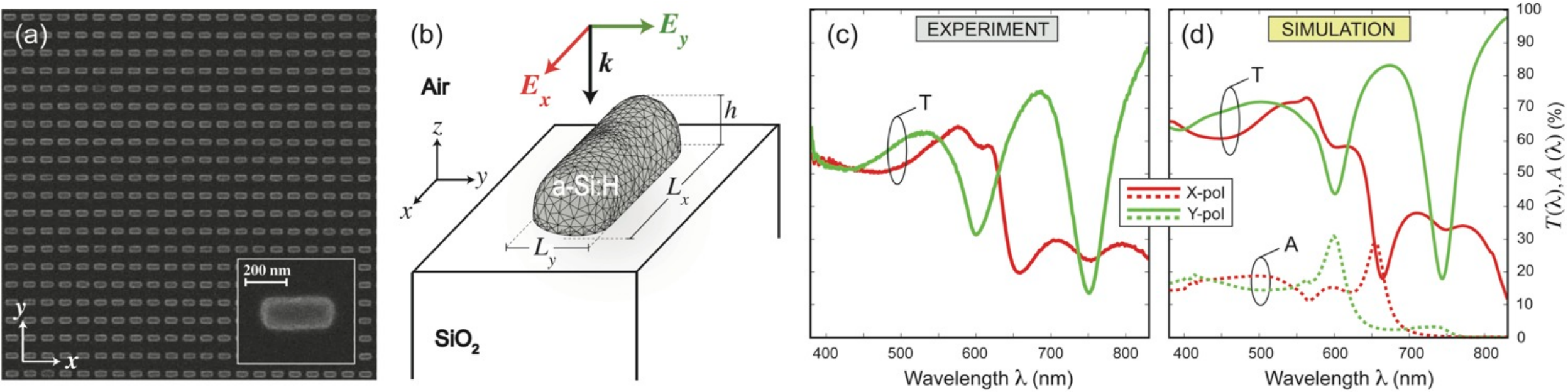}
\caption{(a) SEM image of the dielectric metasurface made of a densely packed 2D array of a-Si:H nanobricks. (b) Sketch of the unit cell considered in the numerical simulations, showing the tetrahedral mesh employed to discretize the a-Si:H meta-atom. The incoming beam is modeled as a plane wave impinging at normal incidence and two different linear polarizations are considered, with the electric field along the x-axis (red) or along the y-axis (green). (c) Measured transmission spectra of the anisotropic metasurface. (d) Simulated transmission (solid lines) and absorption (dotted lines) spectra.}
\label{Fig1}
\end{figure}

A two-dimensional array of a-Si:H nanobricks was manufactured by lithography on a thin a-Si:H film of thickness $d = 175$~nm, which was grown on a silica substrate with plasma-enhanced chemical vapor deposition (see Methods). An SEM image of the a-Si:H metasurface is shown in Fig.~\ref{Fig1}(a). The linear optical response at normal incidence was investigated both experimentally and numerically [cf. Fig.~\ref{Fig1}(b)], and the results are reported in Fig.~\ref{Fig1}(c) and \ref{Fig1}(d). Note that the anisotropic shape of the meta-atom results in a highly anisotropic optical response, with a dominant resonance at 650~nm for the X-polarization ($X$-pol) and two resonances located at 600~nm and 750~nm for the Y-polarization ($Y$-pol). The numerical simulations were performed with full wave finite element software  (CST Microwave Studio) using measured complex permittivity of a-Si:H obtained from the initial 175~nm film of a-Si:H, and are in excellent agreement with the experimental data.

We investigate the transient optical response of the a-Si:H metasurface, by broadband polarization-resolved pump-probe spectroscopy. Importantly, the hydrogenated amorphous silicon (a-Si:H) offers  much faster relaxation of free carriers in comparison to crystalline semiconductors~\cite{Fauchet_JNCS_1992, Baranov_ACSP_2016}, hence the possibility for faster optical modulation. Our experimental setup is based on an amplified Ti:sapphire laser (Coherent, Libra) producing 100~fs pulses at 800~nm wavelength. The sample was excited either by the laser fundamental wavelength at 800~nm or by its second harmonic at 400~nm. The probe pulse was obtained by supercontinuum generation, starting from a fraction of the laser beam (see Supporting Information for further details). 

We measure the differential transmission ($\Delta T/T$), defined as the difference between the transmitted light spectra at normal incidence with and without the pump, normalized to the transmitted light spectrum without the pump.

\begin{figure}[t]
\includegraphics[width=8cm]{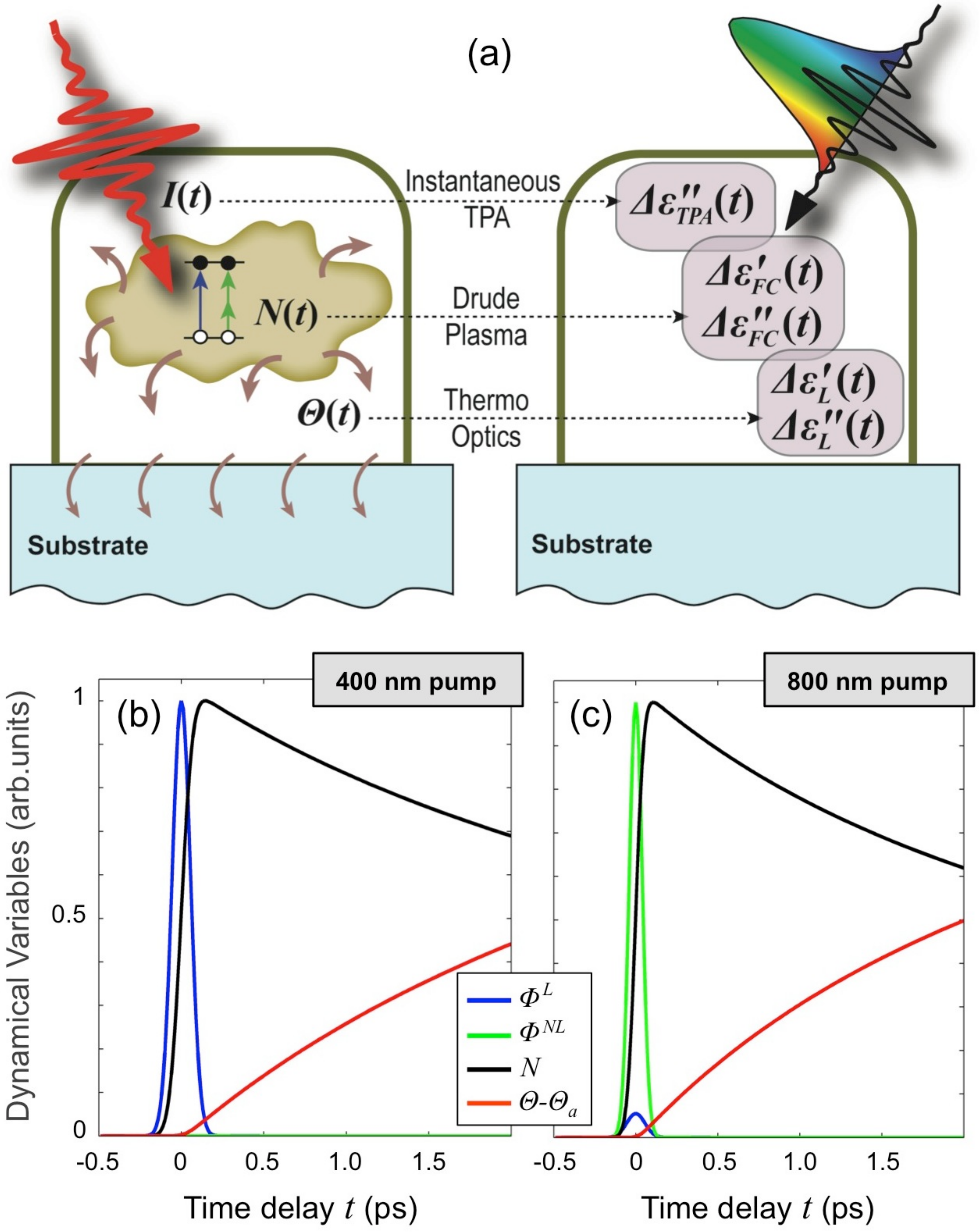}
\caption{(a) Illustration of the nonlinear optical processes in the a-Si:H meta-atom. 
The interaction with a monochromatic pump beam (left panel) gives rise to a modification of the a-Si:H permittivity experienced by the broad-band probe pulse (right panel), according to three different physical mechanisms: instantaneous TPA, Drude plasma response and thermo-optical effect. (b)-(c) Dynamics of the linear $\Phi^{L}(t)$ and nonlinear $\Phi^{NL}(t)$ drives, and of the solution, $N(t)$ and $\Theta(t)$, of the rate equation model governing the nonlinear optical processes. Results for Y-pol pump wavelength excitation at (b) 400~nm and (c) 800~nm are compared. }
\label{Fig2}
\end{figure}

We finally develop a theoretical model for the optical nonlinearity. When a pump pulse of intensity $I(t)$ impinges on the a-Si:H metasurface, free-carriers (electrons in the conduction band and holes in the valence band) are generated, at a rate $\Phi(t)$ per unit volume, by either linear or nonlinear absorption processes, as sketched in Fig.~\ref{Fig2}(a) (left panel). Given a-Si:H is an indirect band gap semiconductor with an energy gap of about $1.7$~eV, the free carrier volume density $N$ relaxes non-radiatively by means of a first-order trap-assisted process, with characteristic time $\tau_{tr} \simeq 30$~ps\cite{Shcherbakov_NL_2015}, or via second-order bimolecular recombination at a rate $\gamma = 2.3\times10^{-8}$~cm$^3$/s~\cite{Fauchet_JOSAB_1989, Esser_PRB_1990}. In order to conserve the energy, each relaxation event occurs alongside phonon generation, which contributes to lattice heating with an energy equal to that of the electron-hole pair~\cite{Fauchet_JNCS_1992}. This causes an increase of the lattice temperature $\Theta$ with respect to the ambient temperature $\Theta_a$. The dynamical properties of the a-Si:H metasurface are thus governed by three variables, the pump pulse incident intensity $I(t)$, the free-carrier density $N(t)$, and the lattice temperature $\Theta(t)$. Each of these variables then  presides over a different mechanism responsible for the pump-induced variation $\Delta\epsilon$ of the complex permittivity experienced by a weak probe pulse of wavelength $\lambda$, arriving on the metasurface at a time delay $t$ with respect to the pump. This is illustrated in the right panel of Fig.~2(a): the intensity $I(t)$ translates into a purely imaginary instantaneous $\Delta\epsilon_{TPA}$ via TPA; the free-carrier density $N(t)$ is responsible for a transient Drude plasma permittivity $\Delta\epsilon_{FC}$, having both real and imaginary parts; finally, the lattice temperature $\Theta(t)$ induces a thermo-optic modulation $\Delta\epsilon_{L}$. Note that the dominant free carrier generation mechanism is linear absorption above the band gap and nonlinear absorption below the band gap. Our three variables are also coupled together by a system of rate equations (see Methods), whose typical evolutions are illustrated in Fig.~2(b) and 2(c) for pump wavelengths above and below the band gap, 400~nm and 800~nm, respectively. One observes, following linear (nonlinear) free carrier generation, their recombination on the picosecond timescale which leads to an increase of the lattice temperature. With the dynamic transient permittivity $\Delta \epsilon$ at hand, the transient optical transmission spectrum of the metasurface as a function of both $\lambda$ and $t$ can be calculated as a perturbation of the previous full wave numerical simulations of the linear system (see Methods).\\

\begin{figure}[t]
\includegraphics[width=16cm]{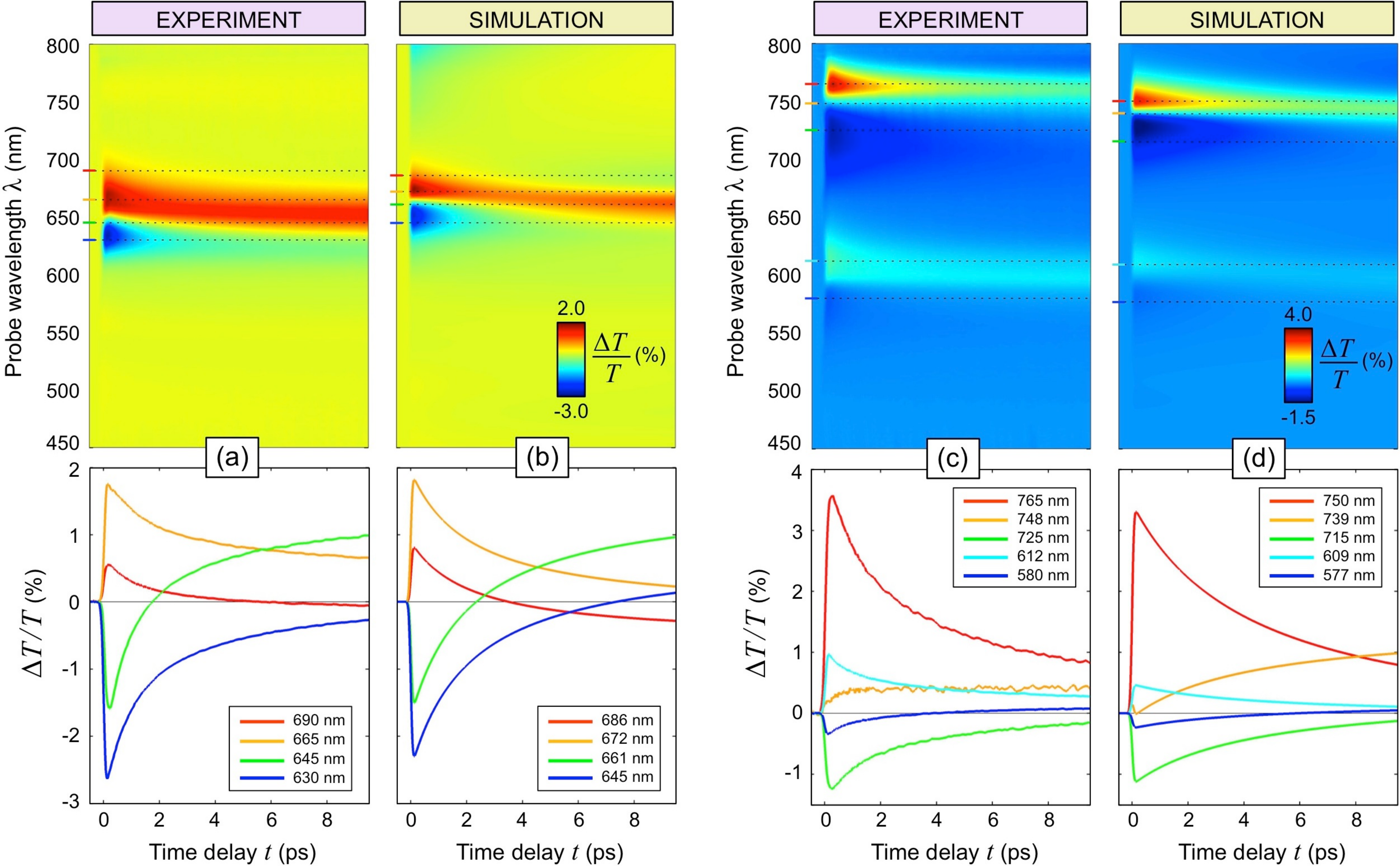}
\caption{Polarization-resolved relative differential transmission under Y-pol pumping with a Gaussian pulse of duration $\tau_p \simeq 110$~fs and fluence $F\sim 0.1\;$mJ/cm$^2$ at 400~nm wavelength. (a) Measurement ($F=105\;\mu$J/cm$^2$) versus (b) simulation ($F=121\;\mu$J/cm$^2$), for X-pol probe. (c) Measurement ($F=83\;\mu$J/cm$^2$) versus (d) simulation ($F=59\;\mu$W/cm$^2$), for Y-pol probe. Top panels show the $\Delta T/T$ maps as a function of time delay $t$ and probe wavelength $\lambda$. Bottom panels show map cross-sections at some selected wavelengths, corresponding to the dot lines in the top panels.}
\label{Fig3}
\end{figure}

The experimental $\Delta T/T$ maps under a Y-pol pump at 400 nm wavelength, and the dynamics at selected probe wavelengths, are reported in Figure~\ref{Fig3}(a) and \ref{Fig3}(c) for X-pol and Y-pol probe, respectively. Following a pulse-width limited build-up, the $\Delta T/T$ signal decays on the timescale of few ps, which is much longer than the pulse duration ($\sim 100$~fs). The initial $\Delta T/T$ spectra are dominated by blue-shifts of the three Mie-like resonances observed in the linear optical response [cf. Figure~\ref{Fig1}(c)], but the scenario then evolves with time. Eventually there can be a sign reversal of $\Delta T/T$, corresponding to a red-shift at long time delays, such as is seen in the dynamics of the Y-pol probe  at 645~nm wavelength [green curve in the bottom panel of Fig.~\ref{Fig3}(a)]. This is a clear indication that the signal is now not due to TPA but rather to the free carriers and hot lattice contributions. However, for different wavelengths, such as the orange curves in the bottom panels of Fig.~\ref{Fig3}(c)-(d), the signal can instead monotonically increase from zero towards a long-lasting plateau within few ps. 
All these features are further accurately reproduced by our model, as detailed by the simulated maps and temporal cross-sections of Figs. 3(b) and 3(d). 
It is worth recognizing that the anisotropy of this transient behavior, such seen in Fig.~\ref{Fig3}(a,b) vs (c,d), can generally allow one to tune the angle of polarization in transmission. 
By utilizing slightly detuned resonances between axes, the X- and Y-pols can experience opposite changes to absolute transmission, and thereby provide mutually constructive polarization rotation.
A similar conclusion could also be made for reflection, given absorption of the probe is negligible.

\begin{figure}[t]
\includegraphics[width=16.5cm]{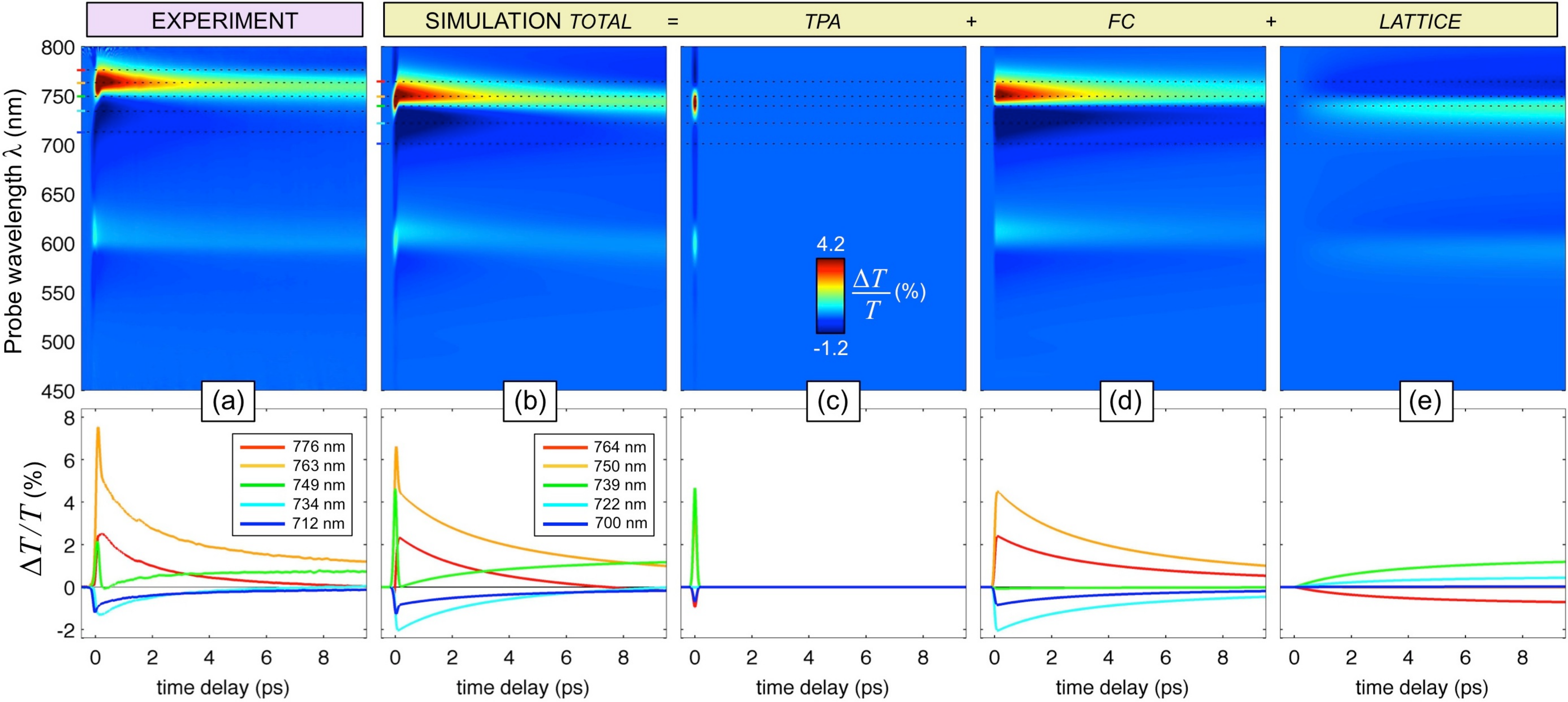}
\caption{Relative differential transmission under Y-pol pumping at 800 nm and Y-pol broad-band probing: (a) Measurement ($F=0.8$~mJ/cm$^2$) versus (b) Simulation ($F=1$~mJ/cm$^2$). Panels (c)-(e) show the three different contributions to the total simulated map of panel (b) given by the instantaneous TPA, the free-carriers, and the lattice. Top panels show the $\Delta T/T$ maps as a function of time delay $t$ and probe wavelength $\lambda$. Bottom panels show map cross-sections at selected wavelengths, corresponding to the dotted lines in the top panels.}
\label{Fig4}
\end{figure}

A different scenario is then observed when pumping in the near infrared. The experimental $\Delta T/T$ map for a Y-pol probe under Y-pol pumping at 800~nm is shown in the top panel of Fig.~\ref{Fig4}(a), together with time traces at selected probe wavelengths (bottom panel). The instantaneous contribution to the transient optical response due to TPA is now very prominent, such as seen in the orange and green traces in the bottom panel of Fig.~\ref{Fig4}(a). This behavior is accurately reproduced by the model, as seen in Fig.~\ref{Fig4}(b), and is observed also in the $\Delta T/T$ map for a X-pol probe, not shown here (see Supporting Information).

No dependence on pump polarization is observed in the transient optical response of the metasurface (as detailed in the Supporting Information), apart from a uniform change in the absolute value of the signal, corresponding to the difference in the anisotropic linear absorption of the pump pulse [cf. dot lines in Fig.~1(d)]. Hence only the two probe polarizations are relevant. The presented theoretical model is able to quantitatively reproduce the complete transient response for both probe polarizations across our broad spectrum, while employing only {\sl two} fixed fitting parameters: (i) the effective TPA coefficient $\beta^{eff}_{TPA}$, and (ii) the $\kappa$ parameter of the imaginary thermo-optic coefficient (defined in the Methods). A fitting procedure retrieves $\beta^{eff}_{TPA} = 0.15$~cm/MW, which is about 3 times higher than the value reported in a-Si:H thin films of comparable thickness, suggesting that the nanostructuring enhances the nonlinear response despite the reduction of the filling factor of the nonlinear medium. The retrieved thermo-optic coefficient was found to be $\kappa=80$~K$^{-1}$~cm$^{-1}$. Considering the substantial dispersion of values reported in the literature, depending on wavelength (about 2 orders of magnitude increase from 750 nm to 650 nm) and on the exact composition of the a-Si:H glass, the value retrieved by our fit is in line with expectations (see e.g. Refs.\citenum{Kovalev_JAP_1996, Poruba_JNCS_2004} and references therein).

Given this theoretical model is able to correctly reproduce experiment, it can now be exploited to elucidate the origin of the spectral and temporal features observed in the experimental $\Delta T/T$ maps. We separate the contributions from each of the three different nonlinear mechanisms taking place under pumping in the near infrared (800 nm), i.e.~the instantaneous TPA [Fig.~4(c)], the Drude plasma response from optically generated free-carriers [Fig.~4(d)], and the thermooptic effect arising from lattice heating [Fig.~4(e)]. This decomposition confirms that the observed slower processes are indeed caused by free-carriers and lattice heating. However free-carriers induce a blue shift of the resonances [Fig.~4(d)], whereas lattice heating is associated with a red shift [Fig.~4(e)]. These two mechanisms can thus partially compensate each other within an individual resonance until the free carriers relaxation is completed. On the contrary, the instantaneous TPA [Fig.~4(c)] causes an increase of transmission at around the peak of the  resonances (600 nm and 750 nm) and a decrease at the sides of these peaks, meaning that TPA results in an instantaneous broadening of the resonances. This explains the peculiar dynamics observed experimentally at around 749~nm [green curve in the bottom panel of Fig.~4(a)] and in the simulations at around 739~nm [the 10~nm blue shift is due to the small shift in the linear spectra of Fig.~1(d) and Fig.~1(c)], where the ultrafast initial peak is followed by the build-up, on the picosecond timescale, of a long-living plateau. This behavior is due to a complete suppression of the contribution arising from the free-carriers, providing zero $\Delta T/T$ at this wavelength, as detailed by the green curve in Fig.~4(d). Similarly, the orange traces and the blue traces in the bottom panels of Fig.~4 correspond to signal wavelengths where the lattice contribution is suppressed in the total $\Delta T/T$, having zero value in the disentangled traces of Fig.~4(e).

\begin{figure}[!ht]
\includegraphics[width=7.8cm]{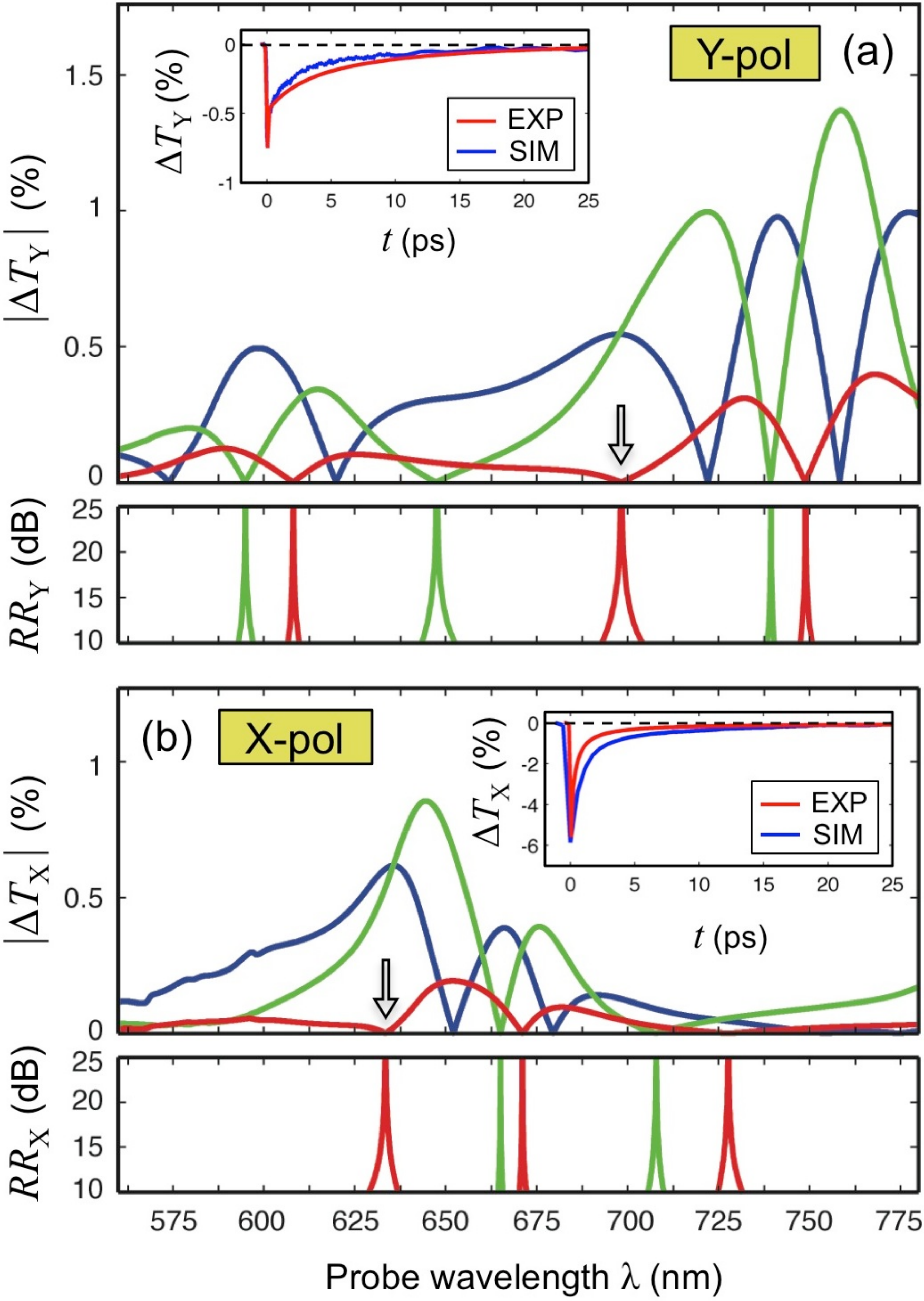}
\caption{High speed modulation windows. (a) Differential transmission spectra (in modulus) for Y-pol probe (pump fluence $F=0.8~\text{mJ/cm}^2$), arising from: the instantaneous TPA at $t = 0$~ps (blue), the free-carriers at $t \simeq 0.1$~ps (green), and the lattice at $t \simeq 20$~ps (red). Bottom panel shows the modulation windows with high rejection of the free-carriers (green) and lattice (red) contributions. Inset shows the experimental $\Delta T_Y (t)$ at 713 nm on a time scale of 25 ps, evaluated at 713~nm, compared with the experimental $\Delta T_Y (t)$ at 700 nm (operating band pointed out by arrow). (b) Same as (a) but for X-pol probe and pump fluence $F = 1.1$~mJ/cm$^2$. Inset show the experimental and simulated $\Delta T_X (t)$ on a time scale of 25~ps, evaluated at around 630~nm (arrow in the main graph) but with $F = 2$~mJ/cm$^2$.}
\label{Fig5}
\end{figure}

The suppression of the contribution from the slowest processes taking place in the transient nonlinear optical response of a-Si:H metasurfaces is of major relevance in view of ultrafast all-optical switching applications. The disentanglement procedure detailed above can be utilized to identify all the wavelength ranges, or operating bands, where such suppression is obtained.
Figure 5 shows the absolute value of the simulated differential transmission $|\Delta T|$ as a function of the probe wavelength [Y-pol probe in panel (a) and X-pol probe in panel (b)], arising from the three different contributions to the optical nonlinearity, each one evaluated at the time delay where the corresponding dynamical variable, either $I(t)$, $N(t)$ or $\Theta(t)$, achieves its maximum [cf. Fig.~2]. Thanks to the peculiar differences between the three physical mechanisms of the all-optical modulation pointed out above, the maximum of the instantaneous $|\Delta T|$ due to TPA (blue curve) is achieved close to those wavelengths where the non-instantaneous contributions, either from the free-carriers (green) or from the lattice (red), approach a negligible value and eventually nullify. To provide a quantitative estimation of this effect, we introduce a figure of merit defined as the polarization-dependent {\it rejection ratio} 
$RR_{X(Y)} = 10\log_{10}(|\Delta T^{fast}_{X(Y)}| / |\Delta T^{slow}_{X(Y)}|)$, 
where  $\Delta T^{fast}$ is the differential transmission due to instantaneous TPA at zero time delay (blue curves in Fig.~5), and $\Delta T^{slow}$ is the differential transmission due to either the free-carriers (at 0.1 ps time delay) or to the lattice heating (at about 20 ps time delay, when the free carrier dynamics are exhausted).  

By plotting the values of $RR_{Y(X)}$ exceeding a given threshold, chosen here at 10~dB, it is possible to identify suitable operation windows where the signal modulation is almost unaffected by the slow dynamics of the system, being due to either the free-carrier Drude response [green curves in the bottom panels of Fig.~\ref{Fig5}(a) and \ref{Fig5}(b)] or due to the thermo-optic effect related to the lattice heating [red curves in the bottom panel of Fig.~\ref{Fig5}(a) and \ref{Fig5}(b)]. Most interestingly, the latter operating bands provide a full return to zero differential signal upon relaxation of the optically generated free-carriers. This is confirmed by the experimental $\Delta T$ traces reported in the inset of Fig.~\ref{Fig5} for the operating bands pointed out by arrows in the main graphs. It is worth recognizing that this is possible despite the fact that the thermal contribution is very long-lasting, taking place on the nanosecond time scale. 
Among these different operating bands with a full-return to zero, the one located at around 700~nm for Y-pol probe is particularly interesting for two reasons: 
(i) the operating band is comparatively broad in-between the resonances [cf. Fig.~\ref{Fig1}(c)-(d)];
(ii) the linear transmission is much higher, implying larger absolute modulation of transmission from TPA, while also enabling the possibility for cascaded operation through consecutive surfaces, due to low reflection losses.
The dynamics of this operating band can then be elucidated by looking at the different transmission modulation mechanisms illustrated in Fig.~4. 
In-between the two Mie-like resonances at 700~nm, the contributions from instantaneous TPA [Fig.~4(c)] superimpose constructively, whereas the contributions from the lattice heating [Fig.~4(e)] superimpose destructively.

We have chosen to keep our study in a perturbative regime where the maximum $|\Delta T|$ is around 1\% [cf. main graphs in Fig.~\ref{Fig5}], however $\Delta T$ can  easily  be increased to the order of 10\% by increasing the pump fluence from 1.1 to 2~mJ/cm$^2$, as demonstrated in the inset of Fig.~\ref{Fig5}(b). 
This was performed when pumping in the near infrared, where the TPA nonlinearity dominates both the instantaneous contribution and as the source of nonlinear free-carrier generation [cf.~Fig.~\ref{Fig2}(c)]. Note that the increase of fluence accelerates the recovery of the signal due to a faster relaxation of the free-carriers induced by a higher bimolecular recombination rate (which is nonlinear in the free-carrier concentration). 
Despite our model being perturbative, a preliminary estimation of the pump fluence required to achieve a full modulation of the transmittance (i.e.~$|\Delta T|\sim|T|$), combined with a sizable transmission in the linear regime is in the order of 3-4 mJ/cm$^2$. This value is compatible with the damage threshold of our metasurface, according to the estimation given in Ref.~\citenum{Shcherbakov_NL_2015} and supported further by the lower $\beta^{eff}_{TPA}$ (and thus lower free-carrier generation and thermal load) of the present configuration.	

In light of the results presented above, we can now outline two further developments for nonlinear anisotropic a-Si:H metasurfaces.\\

1. When the probe is linearly polarized at an angle $\alpha$ with respect to the X-axis, it will experience a modulation given by \mbox{$\Delta T_{\alpha} =\cos^2(\alpha)\Delta T_X + \sin^2(\alpha)\Delta T_Y$}. 
Using our theoretical model (see Methods), the free carrier and lattice heating contributions to $\Delta T_X$ and $\Delta T_Y$ are respectively then linearly proportional to same free carrier density $N$, or lattice temperature $\Theta$.
Subsequently, at any probe wavelength where $\Delta T_X$ and $\Delta T_Y$ have opposite sign, there is guaranteed to be a polarization angle $\alpha$ where the free carrier or lattice heating contribution toward $\Delta T_{\alpha}$ can be made precisely zero, independent of the respective $N$ or $\Theta$.  
This suggests that the high speed modulation windows can be easily tuned by simply rotating the metasurface. 
One could even design an optimized metasurface where, for a particular value of $\alpha$, simultaneous suppression of contributions from both lattice heating and free-carriers is made possible, meaning the recovery to zero of the differential signal ceases to be limited by material response.\\

2. The nonlinear anisotropy of the nanobrick resonators should provide modulation of the phase and amplitude mismatch between the X-pol and Y-pol field components. 
Modulated amplitude mismatch then provides polarization rotation, while phase mismatch provides waveplate transformations.  
This suggests that such a metasurface could operate as an ultrafast new type of all-optical, dynamical wave-plate in a flat-optics configuration.

In conclusion, the presented broadband polarization-resolved pump-probe experiments have revealed a complex scenario for the transient optical response of anisotropic a-Si:H metasurfaces excited by intense femtosecond laser pulses. We have introduced a quantitative model for the observed optical nonlinearity spanning the whole visible spectrum, and validated by the experimental data. This allowed us to disentangle the different physical mechanisms presiding over the all-optical modulation capability of the a-Si:H metasurface. It was found that, despite of the onset of dynamical processes in the a-Si:H material that included very slow thermal effects, a sizable modulation of light transmittance with a full recovery to zero within about 20~ps is achievable in a range of operation windows. Furthermore, the observed ultrafast dynamics can be multiplexed in polarization due to the anisotropy of the metasurface. Our results hence pave the way to the engineering of novel all-dielectric nonlinear metamaterials based on a-Si:H nanostructures, enabling a next generation of ultra-fast all-optical nanophotonic devices, including optical switches and polarization rotators.

\section{Methods}

\subsection{Sample fabrication}
Arrays of silicon nanobricks were fabricated by electron beam lithography on a polycrystalline silicon film grown on a glass substrate via PECVD technique. The substrate was coated with ZEP (a positive-tone electron-beam resist) and baked at 180 C for 120 s. Patterns of silicon bricks were then defined by an electron beam exposure, followed by a development procedure. Subsequently, a 10 nm thick Cr film was deposited by thermal evaporation on the substrate, followed by lift-off. The structures were then transferred to the silicon substrates via a reactive ion etch using the Cr bar nanostructures as etch masks. The residual Cr was then removed via wet etching to obtain the pure Si nanobricks. 

\subsection{Nonlinear model of a-Si:H metasurfaces}
The optically induced dynamical processes taking place in the a-Si:H metaatoms are quantitatively modeled by the following rate equations:
\begin{eqnarray}
\dot{N}(t) &=&-\left(\gamma N(t) + \frac{1}{\tau_{tr}}\right)N(t)+\Phi(t), \label{Eq_2TM_N} \\
C \dot{\Theta}(t) &=&E_{eh}\left(\gamma N(t) + \frac{1}{\tau_{tr}}\right)N(t), \label{Eq_2TM_T}
\end{eqnarray}
In above equations, $E_{eh}$ is the energy of the electron-hole pair, equal to $h\nu_p$ for linear absorption or $2 h\nu_p$ for TPA, $C = 1.66$~J~K$^{-1}$~cm$^{-3}$ is the a-Si:H volume specific heat which is assumed to be equal to that of silicon\cite{Shcherbakov_NL_2015}, and $\Phi(t)$ is the free-carriers generation rate per unit volume that drives the system. The latter is the sum of two contributions, one from linear absorption and one from nonlinear TPA, respectively given by $\Phi^{L}(t) = 1/(h\nu_p) A_L I(t) S/V$ and $\Phi^{NL}(t) = 1/(2 h\nu_p) A_{NL} I(t) S/V$, being $\nu_p$ the frequency of the pump laser pulse with intensity $I(t)$, $S$ the area of the unit cell of the metasurface, $V$ the volume of the metaatom, and $A_{L} (A_{NL})$ the linear (nonlinear) absorption of the metasurface. The linear absorption $A_{L}$ for both X-pol and Y-pol is retrieved from FEM numerical analysis [cf. Fig.~1(d)]. For the nonlinear absorption we assumed the simple isotropic expression $A_{NL} = 1-\exp[-\beta^{eff}_{TPA} I(t) d]$.

The above equations system is numerically solved for a Gaussian pulse of intensity $I(t) = F/(\tau_p\sqrt{\pi/2}) \exp(-2 t^2/\tau_p^2)$, being $F$ the incident fluence and $\tau_p$ the pulse duration. 

The pump incident intensity $I$ is responsible, via TPA, for an instantaneous and dispersionless (i.e.~$\lambda$ independent) variation of the absorption coefficient $\alpha$ of a-Si:H given by $\Delta\alpha(t) = \beta^{eff}_{TPA}I(t)$. This corresponds to an instantaneous imaginary modulation of the permittivity $\Delta\epsilon_{TPA}(t) = i cn'(\nu_p)/(2\pi\nu_p)\Delta\alpha(t)$, where $c$ is the speed of light in vacuum and $n'(\nu_p)$ is the real part of the refractive index of the unperturbed a-Si:H evaluated at the pump frequency. 

The optically generated free-carriers act as a plasma of density $N$, thus providing a variation $\Delta\epsilon_{FC}(\lambda,t) = \Delta\epsilon'_{FC}(\lambda,t)+i \Delta\epsilon''_{FC}(\lambda,t)$ given by the Drude formulas: 
\begin{eqnarray}
\Delta\epsilon'_{FC}(\lambda,t) &=& -\frac{N(t)e^2}{m^* \epsilon_0 (4 \pi^2 c^2\lambda^{-2}+\tau_d^{-2})},\\
\Delta\epsilon''_{FC}(\lambda,t) &=& -\frac{\lambda\Delta\epsilon'_{FC}(\lambda,t)}{2\pi c \tau_d},
\end{eqnarray}
\noindent where $\epsilon_0$ is the vacuum permittivity, $m = 0.12 m_0$ with $m_0$ the free electron mass, and $\tau_d = 0.8$~fs is the Drude damping time (in agreement with Ref.~\citenum{Shcherbakov_NL_2015}).

Finally, the lattice temperature variation $\Delta \Theta(t) = \Theta(t)-\Theta_a$ gives rise, via termo-optic effect, to a permittivity change $\Delta\epsilon_L(\lambda,t) = \Delta\epsilon'_L(\lambda,t)+i \Delta\epsilon''_L(\lambda,t)$ given by:
\begin{eqnarray}
\Delta\epsilon'_L(\lambda,t) &=& 2\left[n'(\lambda)\eta_1-n''(\lambda)\eta_2\right]\Delta \Theta(t),\\
\Delta\epsilon''_L(\lambda,t) &=& 2\left[n''(\lambda)\eta_1+n'(\lambda)\eta_2\right]\Delta \Theta(t),
\end{eqnarray}
\noindent where $n(\lambda) = n'(\lambda) + i n''(\lambda)$ is the complex refractive index of the unperturbed a-Si:H evaluated at the probe wavelength, and $\eta_1 = d n' / d \Theta$ and $\eta_2 = d n'' / d \Theta = \kappa \lambda / (4\pi)$ are the thermooptic coefficients of a-Si:H. We assumed $\eta_1 = 4.5 \times 10^{-4}$~K$^{-1}$ (in agreement with Refs.~\citenum{Fauchet_JOSAB_1989,Shcherbakov_NL_2015}) and $\kappa = 80$~K$^{-1}$~$cm^{-1}$ (fitted on the pump-probe experimental data).

The total $\Delta\epsilon(\lambda,t) = \Delta\epsilon'(\lambda,t) + i \Delta\epsilon''(\lambda,t)$ arising from the superposition of all the different contributions above detailed, is employed to compute the temporal variation of the transmittance spectrum of the optically excited metasurface, $\Delta T(\lambda,t)$, against the transmittance spectrum $T(\lambda)$ of the unperturbed one. This is done perturbatively according to the formula:
\begin{equation}
\Delta T(\lambda,t) = \psi(\lambda;pol)\Delta \epsilon'(\lambda,t) + \phi(\lambda;pol)\Delta \epsilon''(\lambda,t),
\end{equation}
where the polarization dependent spectral coefficients $\psi(\lambda;pol)$ and $\phi(\lambda;pol)$ are given by numerical computation of, respectively, the derivatives $dT/d\epsilon'$ and $dT/d\epsilon''$, evaluated at the probe wavelength (see SI for further details).

\acknowledgement
The authors acknowledge a support by the Australian Research Council and participation in the Erasmus Mundus NANOPHI project (contract number 2013 5659/002-001). GC acknowledges support from the European Union Horizon 2020 Programme under grant agreement No.~696656. GDV acknowledges support by the Italian MIUR through the PRIN 2015 Grant No. 2015WTW7J3. We thank M.R. Shcherbakov, A.E. Miroshnichenko and L. Carletti for useful discussions. The authors acknowledge the use of the Australian National Fabrication Facility (ANFF), the ACT Node.

\bibliography{NanoBricksBiblio}

\end{document}